\documentclass[aps,prl,amsmath,reprint,superscriptaddress]{revtex4-2}
 
\usepackage{graphicx}
\usepackage{float}
\usepackage{color}
\usepackage{upgreek}
\begin{document}
\title{Electron Tunneling Enhances Thermal Conductance through Metal-Insulator-Semiconductor Junctions}
 
\author{Yizhe Liu}
\affiliation{Tsinghua SIGS, Tsinghua University, Shenzhen 518055, China}
\author{Bo Sun}
\email{sun.bo@sz.tsinghua.edu.cn}
\affiliation{Tsinghua SIGS, Tsinghua University, Shenzhen 518055, China}
\affiliation{Guangdong Provincial Key Laboratory of Thermal Management Engineering $\&$ Materials, Shenzhen 518055, China}

\begin{abstract}
The presence of interfaces in semiconductor devices substantially hinders thermal transport, contributing disproportionately to the overall thermal resistance. However, approaches to enhance interfacial thermal transport remain scarce without changing the interface structure, as the intrinsic electron and phonon properties of constituent materials set an upper limit. Here, we find a new thermal transport pathway, electronic heat tunneling, to enhance interfacial thermal conductance through metal-insulator-semiconductor junctions. By applying photoexcitation or bias voltage, we observe remarkable thermal conductance increases \textit{in operando}, opening a new channel for efficient interfacial heat dissipation. The electron quantum tunneling pathway is parallel to conventional phonon-mediated interfacial thermal transport, and violates the Wiedemann-Franz law since this pathway deviates from the paradigm of diffusive transport. Moreover, we develop a tunneling mismatch model to describe the enhanced thermal conductance, originating from tunneling heat flux. Our Letter demonstrates a previously unexplored heat transport mechanism to enhance thermal conductance, bypassing the need for interface engineering. These findings emphasize the essential need to understand semiconductor thermal properties under realistic operating conditions.
\end{abstract}

%\keywords{}

\maketitle
Thermal management of integrated circuits has become a major challenge as the energy density keeps increasing \cite{76,2}. Interfacial thermal resistance is one of the main sources that hinder thermal transport in electronics, severely limiting their performance and reliability \cite{3,4,5}. Furthermore, advancing semiconductor miniaturization leads to growing interface density, urgently requiring an effective way for interfacial heat dissipation. This issue is particularly emergent in structures incorporating dielectric layers, such as metal-insulator-semiconductor (MIS) devices, where dielectrics typically exhibit low thermal conductivity and reduced interfacial thermal conductance \cite{6}. Therefore, enhancing heat conduction across interfaces is an imperative challenge both scientifically and technologically.

Electrons and phonons act as dominant energy carriers in interfacial thermal transport \cite{3,7}. The thermal conductance is fundamentally constrained by constituent materials, since intrinsic properties of electrons and phonons are fixed, and there is no current way for additional carriers to facilitate heat transfer across interfaces. Even for coherent interfaces with atomic sharpness, materials’ intrinsic properties set an upper limit to thermal conductance \cite{5,7,8}. Conventional efforts to enhance thermal conductance focus mainly on material-based issues, such as engineering interface chemical bonding \cite{9,10,11} and composition \cite{12}. However, their application in the modern industry is challenged by the compatibility with existing fabrication processes. To date, methods to increase interfacial thermal conductance remain limited, and the exploration of strategies to improve interfacial thermal transport is in critical demand.

In this Letter, we significantly enhanced the interfacial thermal conductance of the metal/SiO$_2$/Si interface by introducing a new heat transport channel through electron tunneling across MIS junctions. By applying photoexcitation or bias voltage to the metal-SiO$_2$-Si junction, we opened the channel for electrons to tunnel across the SiO$_2$ dielectric layer. The enhanced thermal conductance, originating from electron quantum tunneling, was further evidenced by experiments under different excitation wavelengths and temperatures. Our measurements on both thermal and charge transport revealed that electron quantum tunneling conducted more heat than expected, demonstrating a striking departure from the Wiedemann-Franz law. Our Letter provides firm evidence for electron tunneling heat flux, opening an effective paradigm for advancing interfacial thermal management without structural engineering.

The experiment was conducted as illustrated in Fig. 1(a). We used a 2 $\upmu$m-thick undoped Si membrane with natural SiO$_2$ formed on both sides (see Supplemental Material \cite{13})\vphantom{\cite{14,21,22,38,68,36,53,71,72,37,55,15,16,17,18,19,20,23,24,25,26,27,28,29,30,75,31,32,1,33,34,35,56,57,37,40,39,58,59,60,61,62,63,64,65,66,67}} and deposited an Al film onto the surface to create the MIS junction. The junction was then transferred onto an MgO substrate, which prevents deformation caused by thermal stress and excessive temperature rise due to continuous-wave laser excitation. For photoexcitation applied to the MIS junction, we directed an incident continuous-wave beam (671 nm) with a focused 1/e$^2$ radius of 102 $\upmu$m from the backside into the Si membrane, to generate electrons and holes with a laterally Gaussian distribution (see Supplemental Material \cite{13}). The optical penetration depth of about 4 $\upmu$m \cite{14} is considerably larger than the 2 $\upmu$m thickness of the Si membrane, so we assume that the photoexcited carriers are distributed almost uniformly through the thickness of the Si membrane. We measured the thermal properties of the MIS junction employing time-domain thermoreflectance (TDTR) \cite{21,22,41}, a noncontact optical technique for measuring the thermal transport properties of diverse materials \cite{42,43,44,73,74}, with more details provided in Supplemental Material \cite{13}. We focused TDTR beams with a 1/e$^2$ radius of 2.8 $\upmu$m, which was 2 orders of magnitude smaller than the excitation laser radius. We concentrated the TDTR beams at the Gaussian center of the excitation beam for a sufficient concentration of photoexcited carriers. We modulated the TDTR pump beam at 10.1 MHz to generate periodic temperature oscillations in the MIS junction, with a thermal penetration depth \textit{d} less than 1.42 $\upmu$m, which was shorter than the Si membrane thickness to make sure that the heat flow was confined within the Si membrane.

\begin{figure}
\includegraphics[width=0.48\textwidth,height=0.52\textwidth]{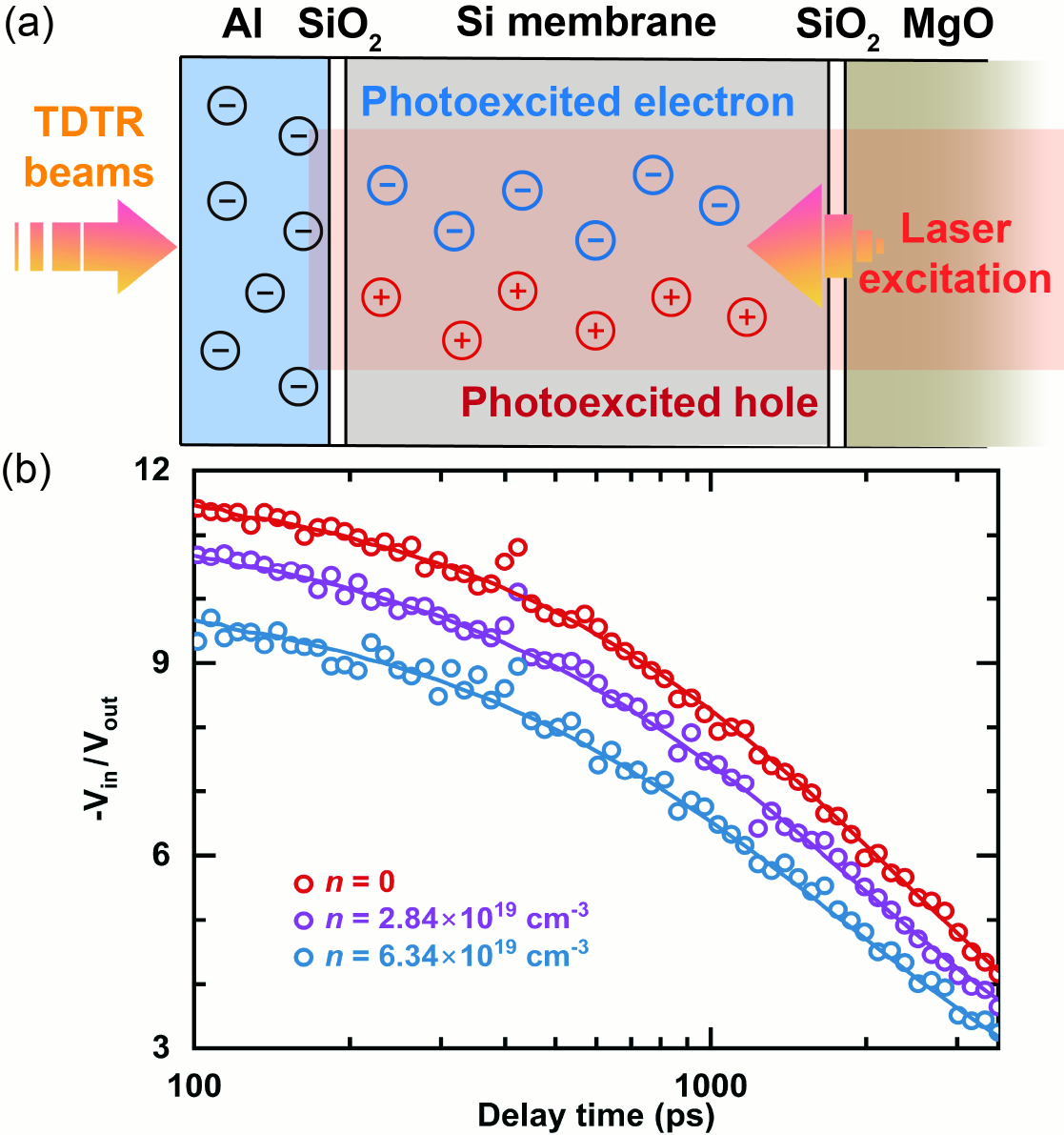}%
\caption{\label{}(a) Schematic illustration depicting the experiment in the Al-SiO$_2$-Si MIS junction under continuous-wave laser excitation. (b) TDTR signals acquired at varying photoexcited electron concentrations. The experimental signals (open circles) were fitted using a multilayered heat diffusion model (solid lines). Red symbols correspond to signals obtained without excitation, while purple and blue symbols correspond to signals under electron concentrations \textit{n} of 2.84$\times$10$^{19}$ cm$^{-3}$ and 6.34$\times$10$^{19}$ cm$^{-3}$, respectively.}
\end{figure}

With continuous-wave laser excitation applied, we found apparent differences in the TDTR signal, as shown in Fig. 1(b). We extracted the thermal conductance of the Al/SiO$_2$/Si interface from TDTR signals obtained in the MIS junction (Fig. 2). The thermal conductance increased continuously with electron concentrations, which exhibits a remarkable enhancement of 23.1$\%$ at the maximum electron concentration (\textit{n} = 6.34$\times$10$^{19}$ cm$^{-3}$), reaching 133 MW m$^{-2}$ K$^{-1}$. This improved interfacial thermal transport due to photoexcited carriers is quite surprising, as direct electron thermal transport should be blocked by the insulator. In addition, electrons’ contribution has been conventionally considered to be in series with the phonon-mediated pathway for interfacial thermal transport across interfaces between metals and nonmetals due to electron-phonon coupling within the Al layer \cite{42}. In this scope, electronic carriers would typically lead to a decrease in interfacial thermal conductance, which is contrary to the observations in our experiments. The temperature rise induced by laser heating had a minimal impact on the measured thermal conductance, as detailed in Supplemental Material \cite{13}.
\begin{figure}
\includegraphics[width=0.44\textwidth,height=0.33\textwidth]{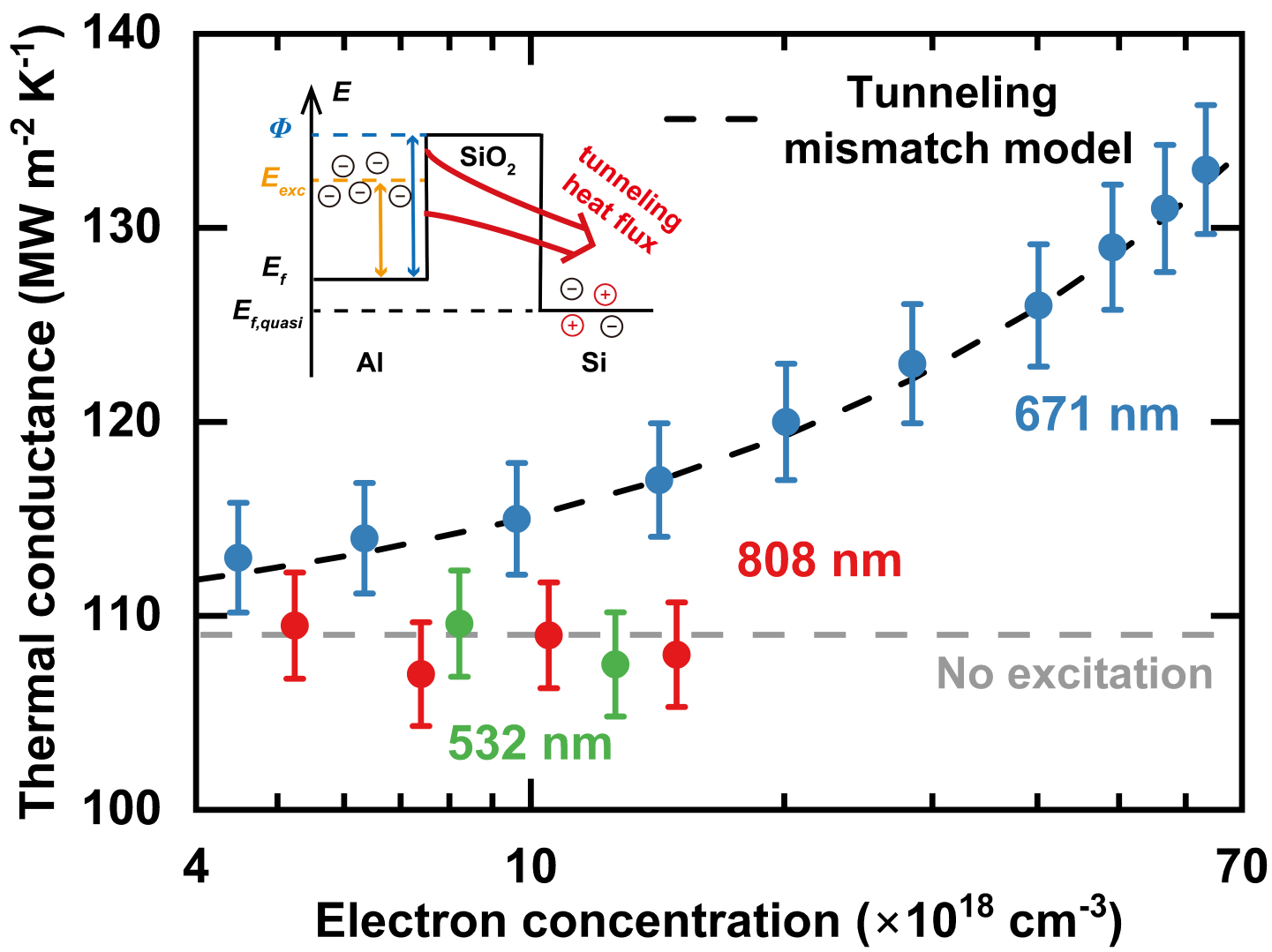}%
\caption{\label{} Measured thermal conductance of the Al/SiO$_2$/Si interface in the MIS junction. Data points corresponding to 532 nm, 671 nm, and 808 nm laser excitation are represented by green, blue, and red circles, respectively. The dashed gray line denotes thermal conductance in the absence of excitation. The inset figure shows band profiles of the MIS junction under continuous-wave excitation, where $\Phi$ denotes the effective barrier height of the SiO$_2$ layer, \textit{E$_f$} is the Fermi energy, and \textit{E}$_{exc}$ represents elevated electron energy by photoexcitation.}
\end{figure}

We then concentrate on the origins of the enhanced thermal conductance of the Al/SiO$_2$/Si interface. Heat conduction across metal/nonmetal interfaces involves four pathways, namely elastic and inelastic phonon-phonon transport, electron-phonon coupling in the metal, and electron-phonon coupling across the interface \cite{5}. The phonon-phonon thermal conductance depends on the phonon density of states and dispersion of the constituent materials on both sides \cite{38}, which is negligibly affected by laser excitation. Given that thermal conductance enhancement correlates with external photoexcitation, electron contribution is the only plausible source to increase thermal conductance. Though electron-phonon interactions across the interface can improve interfacial thermal transport, their contribution is insignificant to the phonon-mediated pathway \cite{45}.  In this case, the elevated thermal conductance points to a buried energy transport pathway across the Al/SiO$_2$/Si interface, driven by electron-electron (or hole) interactions.  Since the thin SiO$_2$ layer impedes direct energy exchange between Al and Si, electronic carriers can conduct heat through quantum tunneling \cite{46}. We should note that energy transfer at metal/insulator/semiconductor interfaces also occurs via several possible pathways, including near-field radiation \cite{68}, thermionic electron injection\cite{36}, Coulomb interaction\cite{69,70}. However, in our setup, the minimal temperature rise rendered near-field radiation and thermionic electron injection to be negligible (see Supplemental Material [13]). The heat flux generated by Coulomb interaction was insignificant compared to phonon-carrier heat\cite{70}, which cannot account for the magnitude of the increased thermal conductance. Therefore, electron quantum tunneling is the only reason for the observed enhancement in thermal conductance.  
\begin{figure}
\includegraphics[width=0.45\textwidth,height=0.3\textwidth]{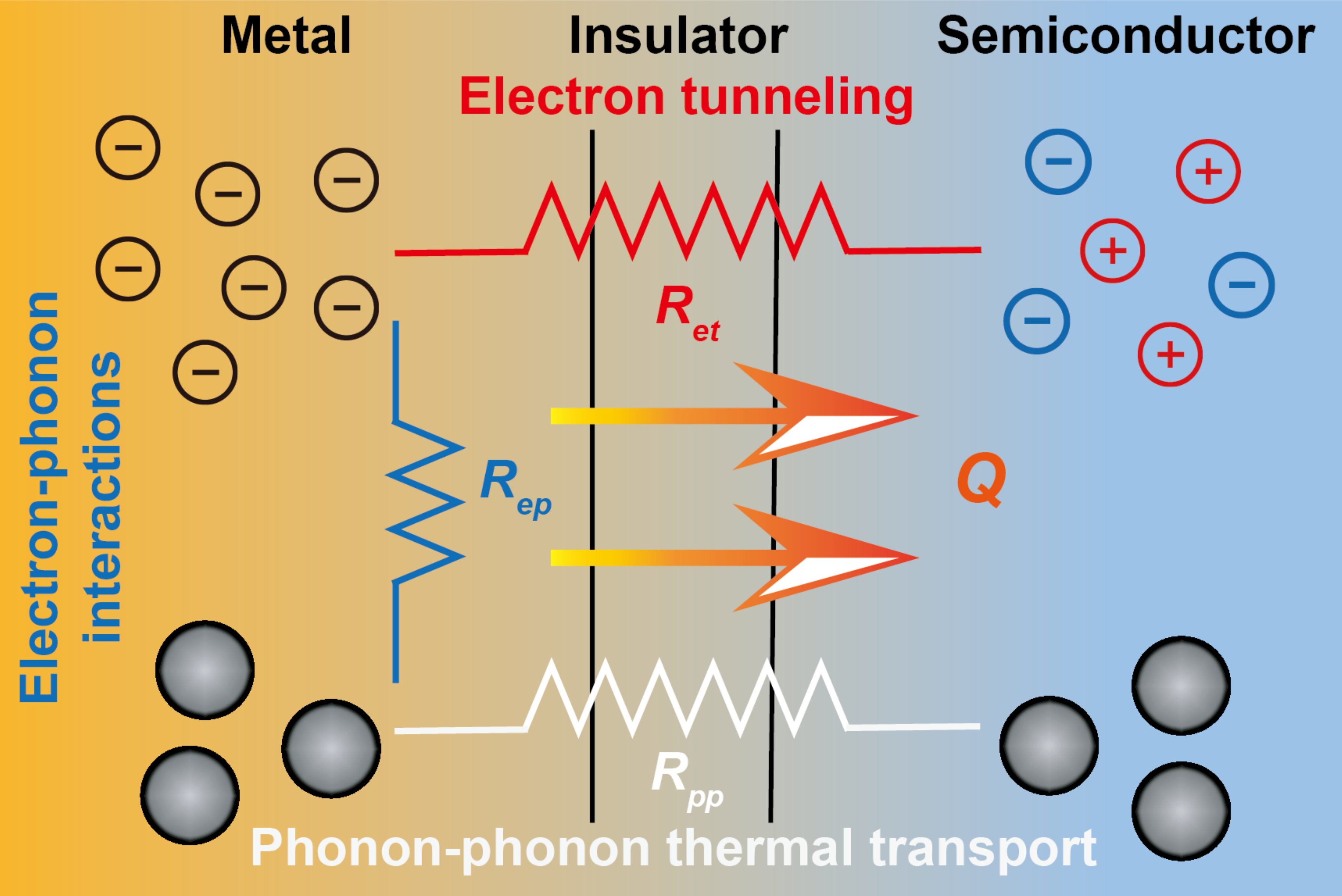}%
\caption{\label{}Primary pathways for heat flow across the metal/insulator/semiconductor interface in our experiment. \textit{R$_{et}$}, \textit{R$_{ep}$}, and  \textit{R$_{pp}$} represent interfacial thermal resistance of electron tunneling, electron-phonon interactions, and phonon-phonon interfacial thermal transport, respectively.}
\end{figure}

The quantum nature of electron tunneling across the interface is further validated by employing excitation with different wavelengths (Fig. 2). Under the excitation of 532 and 808 nm, thermal conductivity of the Si membrane showed a similar decreasing rate to that under the 671 nm laser excitation due to electron-phonon scattering, indicating the presence of sufficient electron concentrations (see Supplemental Material \cite{13}). However, the measured thermal conductance remained nearly unchanged, in sharp contrast to the 671 nm laser excitation results, confirming that the origin of the enhanced thermal conductance is not only photoexcited electronic carriers but also a proper selection of excitation wavelength. For the 671 nm excitation, the corresponding optical penetration depth in Si is more than 2 $\upmu$m, leaving partial photons absorbed by Al near the Al/SiO$_2$/Si interface and elevating the electron energy within Al. Given that the photon energy of the 671 nm laser is insufficient for direct photoemission from Al, electronic carriers transport across the insulating barrier via a quantum tunneling mechanism assisted by photons \cite{47}, rather than a classical transport phenomenon. On the contrary, the 532 nm excitation laser has an optical penetration depth of only 0.8 $\upmu$m in Si, preventing photon injection into Al, and the 808 nm laser offers a relatively lower photon energy, compared to the 671 nm laser. In this case, both wavelengths of excitation fail to increase the interfacial thermal conductance. Our finding provides solid evidence that the enhanced thermal conductance originates from a heat flux driven by electron tunneling, as illustrated in Fig. 3.

To elucidate the electron tunneling heat flux, we provided a quantitative analysis of the enhanced thermal conductance. Although prior research has explored electronic charge tunneling across thin insulators \cite{48,49}, tunneling heat flux has drawn rare attention. Herein, we proposed a tunneling mismatch model to calculate the tunneling thermal conductance \textit{G$_{et}$}, with detailed information provided in Supplemental Material \cite{13}. The model treats electrons and phonons as two separate channels, and the total thermal conductance \textit{G}$_{tot}$ is the sum of the phonon-phonon thermal conductance \textit{G$_p$} and the electron tunneling contribution. As the tunneling probability \textit{P} depends on the effective height and shape of the insulating barrier, which is not readily available, here we apply the tunneling mismatch model to fit the measured thermal conductance enhancements with the parameter \textit{P}. In our experiments, a given excitation elevates the energy of electrons and assists them in tunneling across the insulating barrier. Hence, the tunneling probability follows the Wentzel-Kramers-Brillouin approximation \cite{49}, where detailed information is provided in Supplemental Material \cite{13}. Our measurements align well with calculations by the tunneling mismatch model within experimental uncertainty, and the tunneling probability \textit{P} is extracted as 17.3$\%$ at 300 K, corresponding to the effective barrier height of 1.93 eV (Fig. 2). The analysis of tunneling probability also explains the lack of enhanced thermal conductance under 808 nm laser excitation. When the excitation wavelength increases to 808 nm, the corresponding tunneling probability decreases to 1.56$\%$, leading to a negligible enhancement in thermal conductance within experimental uncertainty. The inset figure of Fig. 2 illustrates photon-assisted electron tunneling in our setup, where the photoexcitation elevates the energy of electrons in Al and assists them in tunneling across the insulating barrier and transporting thermal energy. We then conducted TDTR experiments at 400 K under 671 nm photoexcitation, and the tunneling probability varies negligibly with the temperature (see Supplemental Material \cite{13}). This is expected for electron quantum tunneling, as the tunneling probability is insensitive to temperature \cite{46}. Our TDTR measurements at different temperatures further exclude possible contributions of Coulomb force-generated heat flux, since it is sensitive to the temperature\cite{69}. 
\begin{figure}
\includegraphics[width=0.48\textwidth,height=0.52\textwidth]{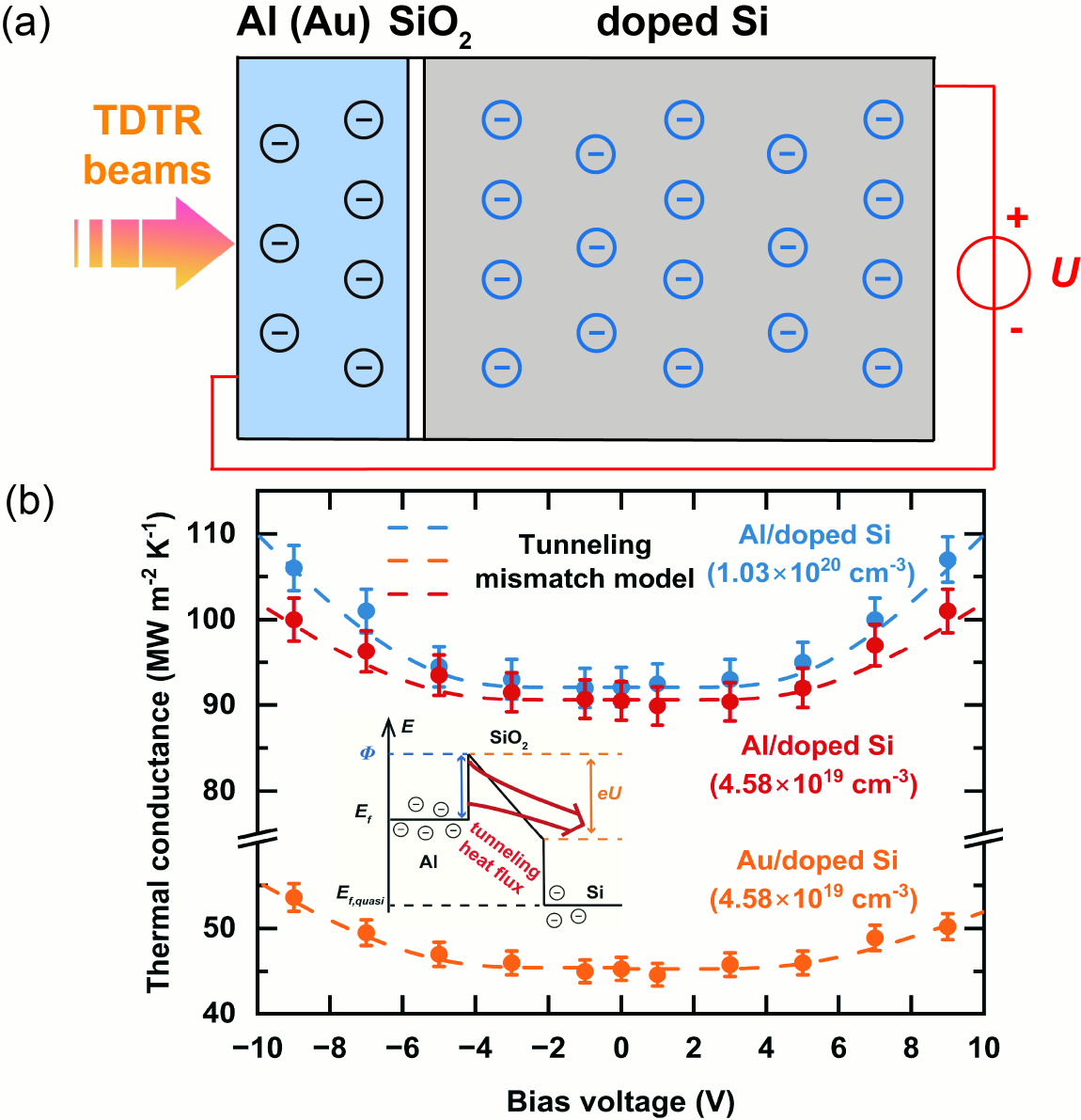}%
\caption{\label{} (a) Schematic illustration depicting the experiment in the Al-thin SiO$_2$-doped Si MIS junction under bias voltages, where the voltage source shows the bias direction. (b) Measured thermal conductance of the Al (or Au)/SiO$_2$/doped Si interface (filled circles) at 300 K, fitted by the tunneling mismatch model (dashed lines). The red plots represent thermal conductance measured in the Al-SiO$_{2}$-doped Si junction (4.58$\times$10$^{19}$ cm$^{-3}$). Blue plots show the measurements in the Al-SiO$_{2}$-doped Si junction (1.03$\times$10$^{20}$ cm$^{-3}$), while orange plots display measurements in the Au-SiO$_{2}$-doped Si junction (4.58$\times$10$^{19}$ cm$^{-3}$). The inset figure shows band profiles of the MIS junction under bias, where \textit{U} stands for the voltage.}
\end{figure}

In practical applications, doped semiconductors with abundant electrons or holes are widely applied in MIS electronics \cite{52}. By applying an external voltage, electronic carriers with adequate energy would tunnel across the insulator layer. This would not only contribute to tunneling charge current but also generate a considerable heat flux, which has been ignored for decades. To mimic the practical work conditions of electronic devices, we built two Al-SiO$_2$-doped Si junctions with different electron concentrations for applying bias voltage, as shown in Fig. 4(a). Before the experiment, electron concentrations of 4.58$\times$10$^{19}$ cm$^{-3}$ and 1.03$\times$10$^{20}$ cm$^{-3}$ for the two doped Si samples were determined through Hall effect measurements (see Supplemental Material \cite{13}). For convenience, we used the electron concentrations to denote the two Al-SiO$_{2}$-doped Si junctions in the following discussions. We found that applying external bias led to an increase in thermal conductance proportional to the bias voltage. Notably, the maximum value of thermal conductance reached 102 MW m$^{-2}$ K$^{-1}$ (4.58$\times$10$^{19}$ cm$^{-3}$) and 108 MW $^{-2}$ K$^{-1}$ (1.03$\times$10$^{20}$ cm$^{-3}$) for the two MIS junctions with enhancements of 13.2$\%$ and 18.8$\%$, respectively, compared to the measured values without bias. The results are displayed in Fig. 4(b). Meanwhile, the thermal conductivity of the doped Si remained unchanged within the experimental uncertainty (see Supplemental Material \cite{13}). In the scenario, applied bias changes the SiO$_2$ barrier shape and opens the tunneling channel for electron-electron energy exchange between Al and Si. 

We then used our tunneling mismatch model to fit experimental results under bias voltages. At low voltages (\textless 3 V), thermal conductance shows a nearly constant value due to lower tunneling probability, while at high voltages, thermal conductance increases apparently owing to remarkably improved tunneling probability. The tunneling probability \textit{P} is characterized by the Fowler-Nordheim model, where the tunneling process is dominated by field emission and the barrier is shaped as a triangle by external voltages \cite{53}. The MIS junction (1.03$\times$10$^{20}$ cm$^{-3}$) exhibited a greater enhancement in thermal conductance compared to the junction (4.58$\times$10$^{19}$ cm$^{-3}$) electron concentration due to a better match in electron density of states and Fermi velocity. The estimated effective barrier height for the Al-SiO$_{2}$-doped Si junction was 2.01 eV, similar to that acquired in the photoexcitation case (1.93 eV).  In addition, we conducted TDTR measurements on the MIS junction at 400 K, where the tunneling probabilities also showed negligible differences compared to those at 300 K (see Supplemental Material \cite{13}). Moreover, we built an Au-SiO$_{2}$-doped Si junction to evaluate the electron tunneling efficiency with different metals. The electron concentration of doped Si was 4.58$\times$10$^{19}$ cm$^{-3}$. The measured  Au/SiO$_{2}$/Si thermal conductance without bias was 44.6 MW m$^{-2}$ K$^{-1}$, which is only half that of the Al/SiO$_{2}$/Si interface due to a greater disparity of Debye temperatures between Au and Si \cite{7}. Under external bias, we achieved a maximum increase of 11.2$\%$ in thermal conductance, which corresponds to the effective barrier height of 2.25 eV according to the tunneling mismatch model. We attribute the increased barrier height for Au to its higher electronegativity and work function, where electrons show a stronger affinity than those in Al. Our demonstration further validated the presence of interfacial thermal transport mediated by electron tunneling in practical MIS devices, breaking the limitations imposed by material-intrinsic thermal conductance. 

We then performed further TDTR measurements under reverse bias on the three MIS junctions, and the results also revealed an enhancement in thermal conductance. This observation reinforces that the enhanced conductance originates from a quantum tunneling process, not from electric-field-driven electron drift. According to our tunneling mismatch model, an asymmetry in thermal conductance between forward and reverse bias was expected when the barrier heights at the metal/insulator and insulator/semiconductor interfaces are dissimilar. The Au-SiO$_{2}$-doped Si junction behaved as predicted, displaying a more pronounced conductance enhancement under reverse bias. Conversely, the Al-SiO$_{2}$-doped Si junctions exhibited largely symmetric behavior with respect to bias polarity. This symmetry is attributed to the comparable barrier heights of the Si-SiO$_{2}$ and Al-SiO$_{2}$ interfaces \cite{71,72}, as the minor difference between them likely fell within the experimental uncertainty of our TDTR measurements.

For deeper insights into tunneling heat flux, we conducted measurements on tunneling charge current of the three MIS junctions, and the results are provided in Supplemental Material  \cite{13}. The current densities in Al-SiO$_{2}$-doped Si junctions were found to be independent of bias polarity, whereas the Au-SiO$_{2}$-doped Si junction displayed a larger tunneling current under reverse bias compared to forward bias (Supplemental Material \cite{13}). Furthermore, at the same applied bias voltage, the Al-SiO$_{2}$-doped Si (1.03$\times$10$^{20}$ cm$^{-3}$) junction sustained the highest current density of the three junctions, while the Au-SiO$_{2}$-doped Si (4.58$\times$10$^{19}$ cm$^{-3}$) junction sustained the lowest. This behavior aligns well with the measured tunneling heat flux and reflects the higher barrier height of Au-SiO$_{2}$ relative to that of Al-SiO$_{2}$ or Si-SiO$_{2}$. We then compared these measured currents with values calculated using the interfacial form of the Wiedemann-Franz law  \cite{37}, which was based on our measured tunneling thermal conductance. This comparison revealed that tunneling electrons have a greater capacity to conduct heat than expected from the Wiedemann-Franz law (see Supplemental Material \cite{13}).  The calculated Lorenz numbers for electron tunneling lie in the range from 3.6\textit{L$_{0}$} to 5.3\textit{L$_{0}$}, where \textit{L$_{0}$} is the Sommerfeld value. This violation of the Wiedemann-Franz law originates from the quantum tunneling pathway, which differs from the conventional diffusive transport.

In summary, this Letter establishes a new pathway for increasing thermal conductance and pioneers the facilitation of interfacial heat exchange by electronic carriers in the quantum tunneling regime. By photoexcitation or applying external voltages, we achieved remarkable enhancements of thermal conductance across Al/SiO$_2$/Si and Au/SiO$_2$/Si interfaces driven by electron tunneling. The quantum tunneling pathway, which differs fundamentally from the diffusive transport regime, leads to a violation of the Wiedemann-Franz law. We believe that electronic heat tunneling is universal in the practical operation of MIS devices. Our Letter enriches experimental explorations into electron tunneling for thermal transport, improving thermal conductance without materials-related phonon engineering. Electron tunneling emerges as a paradigm-shifting approach to overcoming thermal bottlenecks, presenting a promising strategy for managing heat in dense, multilayered electronics.

\begin{acknowledgments}
\textit{Acknowledgments}—We acknowledge the funding support from the National Natural Science Foundation of China (No. 52161145502 and No. 12004211), the Shenzhen Science and Technology Innovation Program (No. RCYX20200714114643187 and No. WDZC20200821100123001), the Tsinghua Shenzhen International Graduate School (No. QD2021008N and No. JC2021008), and the Guangdong Special Support Plan (No. 2023TQ07A273).

\textit{Data availability}—The data that support the findings of this article are openly available \cite{77}. 
\end{acknowledgments}

\bibliographystyle{apsrev4-2}

\end{document}